\documentclass[10pt,letterpaper]{article}
\usepackage{opex3}
\usepackage{color}
\usepackage{graphicx}
\usepackage{epstopdf}
\usepackage{cite}
\usepackage[colorlinks=true,linkcolor=blue,citecolor=blue]{hyperref}
\usepackage{textcomp}
\begin{document}

\title{Towards self-similar propagation in a dispersion tailored and highly nonlinear segmented bandgap fiber at 2.8 $\mu$m}

\author{Piyali Biswas$^1$, Somnath Ghosh$^{1*}$, Abhijit Biswas$^1$, and Bishnu P. Pal$^2$}

\address{$^1$Institute of Radio Physics and Electronics, University of Calcutta, Kolkata, India\\
	$^2$Mahindra \'{E}cole Centrale, School of Natural Sciences, Hyderabad-500043, India}

\email{$^*$somiit@rediffmail.com} 



\begin{abstract}
We numerically demonstrate self-similar propagation of parabolic optical pulses through a highly nonlinear and passive specialty photonic bandgap fiber at 2.8 $\mu$m. In this context, we have proposed a scheme endowed with a rapidly varying, but of nearly-mean-zero longitudinal dispersion and modulated nonlinear profile in order to achieve self-similarity of the formed parabolic pulse propagating over longer distances. To implement the proposed scheme, we have designed a segmented bandgap fiber with suitably tapered counterparts to realize such customized dispersion with chalchogenide glass materials. A self-similar parabolic pulse with full-width-at-half-maxima of 4.12 $p$s and energy of $\sim$ 39 $p$J has been achieved at the output. Along with a linear chirp spanning over the entire pulse duration, 3dB spectral broadening of about 38 nm at the output has been reported. 
\end{abstract}

\ocis{(190.4370) Nonlinear optics, fibers; (320.7110) Ultrafast nonlinear optics; (060.5295) Photonic crystal fibers} 


\section{Introduction}
Nonlinear propagation of optical pulses at the mid-infrared (mid-IR) ranging from 2 to 25 $\mu$m have become a tremendous interesting domain of research in science and technology due to its wide potential applications specially in biomedical surgeries, spectroscopy, chemical/bio-molecular sensing and so on \cite{Ilev2006,Serebryakov:10}. Such applications require efficient mid-IR laser sources and suitable light-guiding structures. Recent developments on mid-IR lasers and availability of fabrication-compatible low-loss nonlinear infrared (IR) glasses have eventually accelerated the study of nonlinear pulse propagation \cite{sanghera,wang,JACE}. Moreover, presence of self-similarity in complicated and nonlinear systems have made it effortless to understand various nonlinear physical phenomena. Generally, dynamics of any nonlinear systems in physics are governed by several similar kind of complicated partial differential equations. In order to have a simplified representation of nonlinear properties, self-similarity is established to reduce the number of degrees of freedom of the governing equation by reduction of symmetry technique, thereby obtaining a self-similar solution of such equations \cite{barenblatt1996scaling}. 

Self-similarity has extensive exploitation in various branches of physics, and also is rapidly emerging in the domain of nonlinear optics. Nonlinear Schr\"{o}dinger equation (NLSE) is the governing equation in nonlinear fiber optics, solution of which in normal group velocity dispersion (GVD) regime is asymptotically self-similar with a temporal parabolic power profile \cite{exactss,hirooka}. Such self-similar parabolic solutions of NLSE has a tremendous capacity of tolerating large nonlinearity without wave breaking \cite{anderson}. Further, increasing demand of ultra-short, high power pulse propagation have attracted considerable attention of researchers towards self-similar propagation. Initially, the ultra-short high power pulses suffered from the deleterious effect of optical wave breaking (OWB) due to accumulation of large nonlinear phase shift during pulse propagation \cite{wavebreak}. Also, investigations have revealed that OWB can be avoided by formation of self-similar parabolic pulses (PP) with linear chirp across its width. To date, a number of works have extensively studied and reported the propagation of apparently high power laser pulses via formation of PPs, both in active and in passive fibers \cite{fermann,kruglov,wise,parm,finot07,Iakushev,laserss,lavdas,jiang,loworder,barh}. Moreover, suitable fiber designs have been proposed for PP formation \cite{dghosh}. However, recent investigations have revealed that PPs that are formed in highly nonlinear and passive fibers are unable to reach steady state regime of propagation due to excessive nonlinearity overshadowing dispersion etc. As a result, the parabolic temporal profile  gradually reshapes itself into a nearly rectangular profile leading to OWB. Here, the formed PPs are mere intermediate transient state of propagation and fail to attain self-similarity \cite{nfactor,boscolo}. Certain nonlinear pulse propagations have very recently been reported in highly nonlinear and dispersion decreasing microstructured optical fibers (MOF), where investigations reveal that formed PPs do not retain their shape over longer propagation length as the large nonlinearity of the chosen MOF broadens the pulse in temporal domain as well as in spectral domain. Moreover, pulse propagation has essentially no dependence on choice of different dispersion profiles in that regime \cite{piyali}. Thus for the applications of high power ultra-short pulses, such dynamics of pulse propagations are not suitable, and require newer scheme/fiber design to handle the specific difficulty of pulse propagation. Lately, efforts have been put forward towards high power self-similar pulse evolution through highly nonlinear fiber lasers (both active and passive), though it has been realized that in practice it is challenging to obtain the best performance of self-similar PP formation in highly nonlinear passive media \cite{chong}. 

In our work, we propose a new scheme based on an unconventional dispersion profile of specialty optical fibers to achieve self-similarity, and demonstrate propagation of self-similar parabolic pulses through a highly nonlinear and passive specialty photonic bandgap fiber (PBG) at the mid-IR wavelength of 2.8 $\mu$m. For this purpose, we have chosen suitable chalcogenide glasses for designing the PBG fiber: $GeAsSe$ as the low index core material and combination of $AsSe/GeAsSe$ as the cladding material. In order to obtain self-similar propagation for the formed PPs, we have adopted a rapidly varying dispersion profile with a mean value close to zero so that effect of dispersion accumulated phase shift on pulse can be reduced significantly. Further, we have designed a segmented highly nonlinear bandgap fiber (HNBF) exploiting the aforesaid scheme. Formation of high power PPs with self-similar propagation over longer lengths of the designed HNBF have been reported. Characteristic temporal and spectral broadening of PPs have also been achieved with proper linear chirp across the pulse width. 

\section{Numerical Model}
Parabolic pulses with linear chirp across its width are formed through interplay between SPM and positive GVD. In highly nonlinear passive fibers, primarily nonlinearity drives the dynamics of such PPs overshadowing dispersion and hence fails to attain self-similarity. In order to investigate self-similar propagation of ultra-short high power PPs in highly nonlinear bandgap fiber (HNBF) taking into account the interplay between nonlinearity and dispersion, we have adopted a numerical methodology. To study the dynamics of such pulse propagation through dispersion and nonlinearity tailored HNBF, we need to solve the NLSE numerically as it well-describes the problem of short pulse propagation. Considering a slowly varying picosecond (ps) pulse envelope $A(z,t)$, NLSE takes the form \cite{book},
\begin{equation}
\frac{\partial A}{\partial z}+\beta_1 \frac{\partial A}{\partial t}+i \frac{\beta_2}{2} \frac{\partial^2 A}{\partial t^2}+ \frac{\alpha}{2} A = i \gamma(\omega_0) |A|^2A, 
\end{equation}
where nonlinear parameter $\gamma$ is defined as,
\begin{equation}
\gamma(\omega_0) = \frac{n_2(\omega_0) \omega_0}{c A_e}.
\end{equation}
$\alpha$ is the loss parameter, $\beta_1$ and $\beta_2$ are the first and second order dispersions respectively, $A_e$ is the effective mode area of the fiber and $n_2$ is the nonlinear coefficient of the medium. The pulse $A(z,t)$ is assumed to move in a retarded reference time frame moving with a group velocity $v_g$,where $T=t-z/v_g$$\equiv$$ t-\beta_1z$ and is normalized such that $|A|^2$ represents the optical power.

In an optical fiber with normal GVD i.e., when value of $\beta_2$ is positive and a hyperbolic dispersion decreasing profile along length of the fiber, the asymptotic solution of NLSE yields a parabolic intensity profile. Under this condition, the propagation of optical pulses is governed by the NLSE of the form,

\begin{equation}
\label{eq:3}
i \frac{\partial A}{\partial z}-\frac{\beta_2}{2} D(z) \frac{\partial^2 A}{\partial T^2}+i\frac{\alpha}{2}A+ \gamma(z)|A|^2A = 0, 
\end{equation}
where $D(z)$ is length dependent dispersion profile and customized according to our proposed scheme, $\beta_2$ (GVD parameter) $>$ $0$ and $\gamma(z)$ is longitudinally varying nonlinear (NL) coefficient. The dispersion and nonlinear parameters i.e. $\beta_2$ and $\gamma$ are calculated along the propagation length of the segmented HNBF at the central wavelength 2.8 $\mu$m. With the use of coordinate transformation $\xi=\int_{0}^{z}D(z')dz'$ and defining a new amplitude $U(\xi,T)=\frac{A(\xi,T)}{\sqrt{D(\xi)}}$, eq. (3) transforms to,

\begin{equation}
\label{eq:4}
i \frac{\partial U}{\partial \xi}-\frac{\beta_2}{2} \frac{\partial^2 U}{\partial T^2}+\frac{\alpha}{2}U+ \gamma(z)|U|^2U = i \frac{\Gamma(\xi)}{2}U, 
\end{equation}
where \begin{equation} 
\label{eq:5}
\Gamma(\xi)=-\frac{1}{D}\frac{dD}{d\xi}=-\frac{1}{D^2}\frac{dD}{dz}
\end{equation}
Equation (\ref{eq:5}) clearly dictates that $D(z)$ is varying and a decreasing function of $z$. Hence decreasing dispersion acts exactly as the varying gain term of a fiber amplifier with normal GVD. Specifically, with the choice of $D(z) = \frac{1}{1+\Gamma_0 z}$ the gain coefficient becomes constant, i.e., $\Gamma$ = $\Gamma_0$. Here, the NLSE with normal and customized GVD, and constant gain coefficient delivers an asymptotic self-similar parabolic solution with a linear-chirp across its width. For subsequent discussion in this paper, it is convenient to use following notations: $L_D$ is the dispersion length, $L_{NL}$ nonlinear length and soliton number $N$,
\begin{equation}
L_D=\frac{T_0^2}{|\beta_2|}, \qquad L_{NL}=\frac{1}{\gamma P_0}, \qquad N=\sqrt{\frac{L_D}{L_{NL}}}
\end{equation}
where, $P_0$ is the peak power of the initial pulse, $T_0$ is the initial pulse duration (half-width at $1/e$ intensity point).
Now, equation (\ref{eq:4}) has been solved with the help of Split-step Fourier Method (SSFM). Though there are a number of numerical approaches, SSFM provides the most easy and fastest way to solve NLSE numerically. Here, pulse propagation is carried out from $z$ to a very small incremental length $\Delta$$z$ in two steps: only dispersion dominated first step and only nonlinearity dominated second step. In this method eq.(\ref{eq:4}) can be written in the form \cite{book},
\begin{equation}
\frac{\partial A}{\partial z}=(\hat D+\hat N)A,
\end{equation}
where $\hat D$ is the differential operator that accounts for the dispersion and losses within the medium, and $\hat N$ is the nonlinear operator that governs the effect of fiber nonlinearities on pulse propagation. 

In order to check the quality of the formed parabolic pulse, we compute the evolution of the misfit parameter $(M)$ between the pulse temporal intensity profile $|U|^2$ and the parabolic fit $|p|^2$ of the same energy \cite{barh}:
\begin{equation}
M^2=\frac{\int[|U|^2-|p|^2]^2dT}{\int|U|^4dT}
\end{equation}
where, the expression for the parabolic fit with peak power $P_p$ and pulse duration $T_p$ is given by:
\begin{equation}
p(T)=\left\{\begin{array}{rcl}{P_p(1-\frac{T^2}{T_p^2})}&\mbox{for}&|T|\leq|T_p|\\0&\mbox{for}&|T|>|T_p|\end{array}\right.
\end{equation}
The smaller values of $M$ show better fit to the targeted parabolic waveform. We have considered $M$$<$0.04 which is sufficient for a pulse to be parabolic. All numerical simulations are carried out in MATLAB\textregistered.

\section{Self-similar pulse propagation}
\subsection{Fiber design based on the proposed scheme}
In order to achieve self-similar propagation, we propose a new method based on a suitable dispersion and nonlinearity tailoring in highly nonlinear media along the propagation length. The scheme is endowed with a rapid variation of dispersion and nonlinearity over the propagation length. Specifically, the longitudinal dispersion profile rapidly varies around a mean which is nearly zero and positive, while the corresponding fiber nonlinearity has been simultaneously modulated around a much higher value. To implement the proposed scheme, we target to design a specialty
\begin{figure}[htbp]
	\centering\includegraphics[width=12cm,trim={0 0.4cm 0 0}]{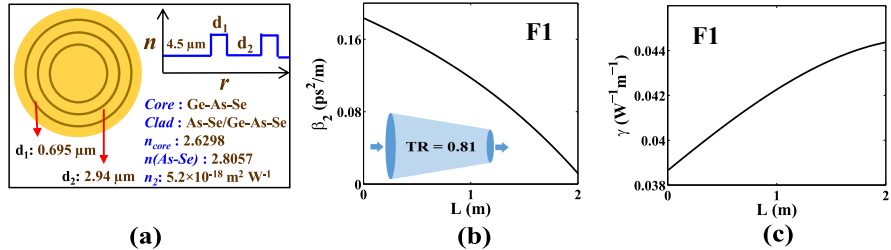}
	\caption{(a) Input cross-section with refractive index profile of the designed photonic bandgap fiber F1; Variation of (b) GVD parameter ($\beta_2$) and (c) nonlinear parameter ($\gamma$) along the length of F1.}
	\label{fig:figure1}
\end{figure}
optical fiber to support self-similar parabolic pulse at 2.8 $\mu$m in the mid-IR wavelength range. The availability of laser source at 2.8 $\mu$m have driven us to design a parabolic pulse source at such longer wavelength. Further, fabrication-compatible chalcogenide glasses at mid-IR are highly nonlinear in nature. To meet the challenges to form self-similar pulses in such fibers, we design two HNBF with a customized dispersion and nonlinear profiles. 
\begin{figure}[htbp]
	\centering\includegraphics[width=12cm,trim={0 0.3cm 0 0}]{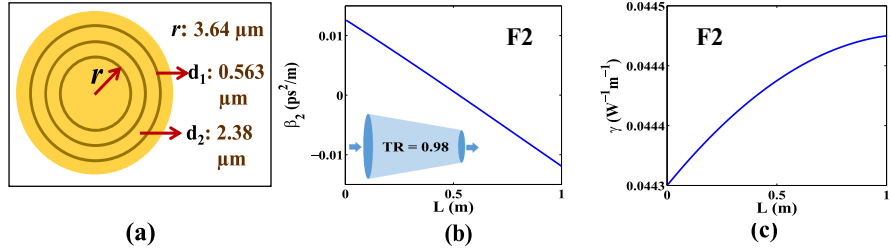}
	\caption{(a) Input cross-section of the 1 m long designed HNBF F2 with its dimensions; Longitudinal variations of (b) GVD parameter ($\beta_2$), and (c) nonlinear parameter ($\gamma$) for F2. }
	\label{fig:figure2}
\end{figure}

Initially, we have designed a dispersion decreasing bandgap fiber (DDBF) for PP formation where light is guided by the customized PBG around the chosen wvelength. To achieve this, only two chalcogenide glass materials have been chosen: (a) $GeAsSe$ having a lower refractive index of 2.6298 \cite{irg24}, and (b) $AsSe$ with higher refractive index of 2.8057 \cite{irg26}, both at 2.8 $\mu$m. The designed HNBF has $GeAsSe$ made core of radius 4.5 $\mu$m. Surrounding the core, there are three successive bi-layers of cladding thereby resulting the desired PBG. Here, a cladding bi-layer consists of a high-indexed $AsSe$ clubbed with a low-indexed $GeAsSe$. The nonlinear refractive index $n_2$ of the core material is very high, $\sim$ 5.2$\times$10$^{-18}$ m$^{-2}$/W at 2.8 $\mu$m. We denote this designed HNBF as F1. The cross-section of F1 is shown in Fig. \ref{fig:figure1}(a). At 2.8 $\mu$m for F1, the value of GVD parameter $\beta_2$ is 0.1835 $p$s$^2$m$^{-1}$ and nonlinear parameter $\gamma$ is 0.0386 W$^{-1}$m$^{-1}$. Further, starting with these parameter values, F1 has been down-tapered with a taper-ratio 0.81 over a length of 2 m. The output end cross-sectional parameters of F1 are (i.e., at 2 m): core-radius $r$ = 3.645 $\mu$m, $d_1$= 0.563 $\mu$m, $d_2$= 2.381 $\mu$m, $\beta_2$= 0.01268 $p$s$^2$m$^{-1}$, and $\gamma$= 0.04434 W$^{-1}$m$^{-1}$. The longitudinal variation of $\beta_2$ and $\gamma$ for F1 is shown in Fig. \ref{fig:figure1}(b) and Fig. \ref{fig:figure1}(c), respectively.

\begin{figure}[htbp]
	\centering\includegraphics[width=12cm,trim={0 0.3cm 0 0}]{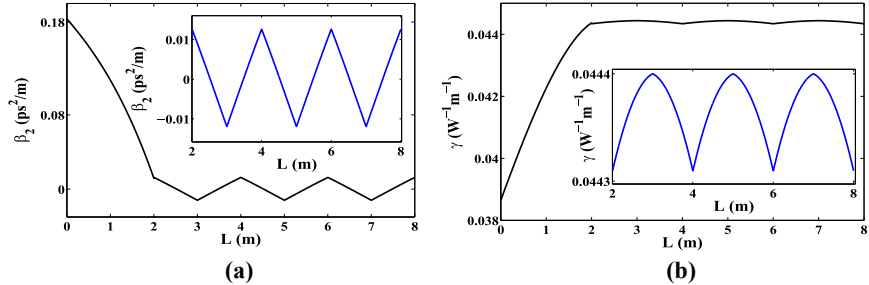}
	\caption{(a) Variation of $\beta_2$ along the total combined length of F1 and F2 with six up-down repetitions in order to form the rapidly varying dispersion profile. Inset shows the $\beta_2$ variation of repeated F2 with length; (b) variation of $\gamma$ along the total length of combined HNBF. $\gamma$ profile over repeated F2 only is shown as inset.}
	\label{fig:figure3}
\end{figure}

The other HNBF, denoted as F2, is 1 m long with same material as mentioned above keeping the input-end cross-section identical to the output-end cross-section of F1 in order to avoid any discontinuity in parameter values. F2 is down tapered with a taper-ratio of 0.98, so that the value of $\beta_2$ starts from positive, crosses zero value and reaches negative dispersion regime. Now this specific dispersion profile of F2 is repeated in such a way over the next few meters that we realize a rapidly varying dispersion profile with a nearly mean-zero $\beta_2$ value. This particular mode of dispersion tailoring in F2 is deliberately exploited in order to customize the dispersive effect of fiber on pulse propagation (along with related $\gamma$ modulations) so as to achieve self-similarity of the formed PP. The cross-section of F2 is shown in Fig \ref{fig:figure2}(a). Moreover, the longitudinal variation of $\beta_2$ and $\gamma$ is depicted in Fig. \ref{fig:figure2}(b) and \ref{fig:figure2}(c), respectively. The entire dispersion profile of the proposed segmented fiber with both F1 and F2 connected together following the proposed scheme are depicted in Fig. 3(a). Similarly, variation of $\gamma$ is shown in Fig. 3(b) for the segmented HNBF over an arbitrarily chosen length of 8 m. 
\subsection{Pulse evolution}
In our study, we use an un-chirped Gaussian shaped initial pulse waveform centred at the wavelength of 2.8 $\mu$m, which may be obtained by using a mode-locked Er$^{3+}$-doped ZBLAN fiber laser as optical source \cite{zblan}. We have considered the peak power of the initial pulse to be 150 W and full-width-at-half-maxima (FWHM) to be 2.0 $p$s, conditioned by the characteristics of the laser source. The initial pulse has been injected at the input end of the segmented HNBF and allowed to propagate down the entire fiber length of 8 m. Investigation reveals that after propagating only 1.5 m from
\begin{figure}[htbp]
	\centering\includegraphics[width=12cm,trim={0 0.3cm 0 0}]{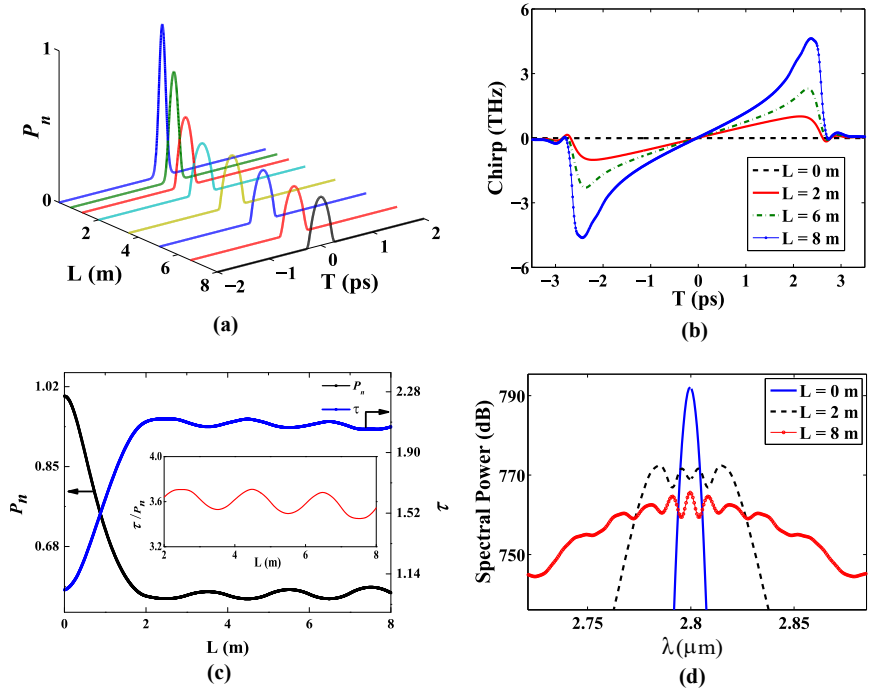}
	\caption{(a) Evolution of self-similar pulses through the entire length of the segmented HNBF; (b) Variation of chirp developed at different HNBF length; (c) Behavior of normalized pulse power ($P_n$) and normalized FWHM ($\tau$) along fiber length with variation of the ratio $\tau$/$P_n$ as inset; (c) Spectral variation at different fiber lengths.}
	\label{fig:figure4}
\end{figure} 
the input end of the HNBF, Gaussian pulse starts approaching towards parabolic intensity profile in time domain. Misfit parameter ($M$) computation for pulse evolution dictates that at 1.55 m from the input end of the HNBF, PP has been evolved with the minimum value of $M$=0.0179. The overall pulse evolution is depicted in Fig. \ref{fig:figure4}(a). Explicit numerical study of the propagation dynamics reveals that once PP is formed in the HNBF, it maintains its shape over rest of the segmented HNBF length. Evolutions of PPs formed at different lengths and their corresponding chirp variation is shown in Fig. \ref{fig:figure4}(b). In Fig. \ref{fig:figure4}(c), we have shown the simultaneous variation of normalized pulse power ($P_n$) and normalized FWHM ($\tau$) along the entire fiber length which clearly exhibit their complementary nature. To confirm the self-similar behavior of PP, we have computed the ratio of $\tau$ to $P_n$ and have shown the variation of $\tau$/$P_n$ over 6 m of HNBF length as inset of Fig. \ref{fig:figure4}(c). Computations reveal that the $\tau$/$P_n$ variation is nearly constant with an average of 3.595 and relative standard deviation (RSD) turns out to be as low as 2.28$\%$. Thus, at the output end of the fiber, a self-similar PP with FWHM 4.12 $p$s and energy $\sim$ 39 $p$J has been observed. Moreover, Spectral broadening of $\sim$ 38 nm at 3dB points has been reported. Spectral changes during pulse propagation are explicitly depicted in Fig. \ref{fig:figure4}(d).

\begin{figure}[htbp]
	\centering\includegraphics[width=12cm,trim={0 0.3cm 0 0}]{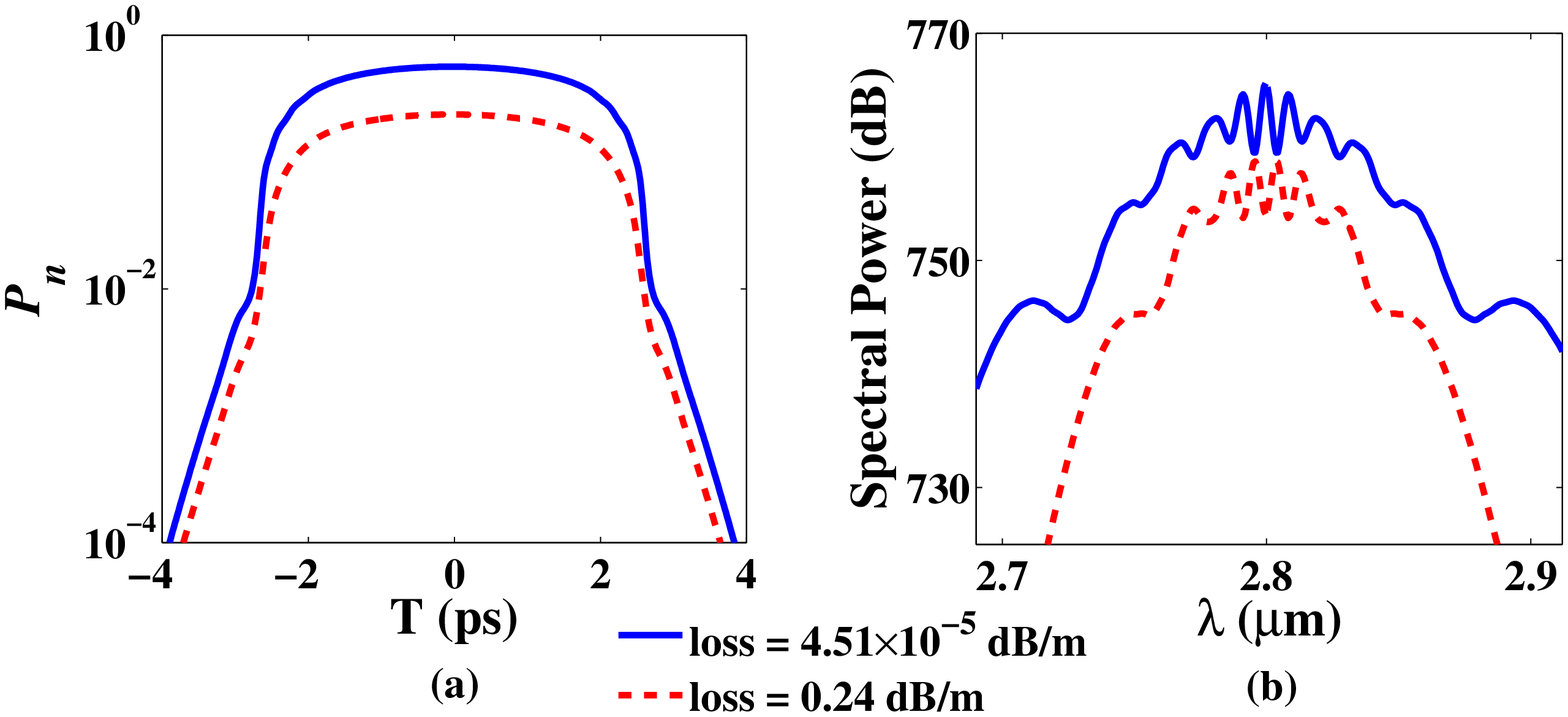}
	\caption{Effect of loss on (a) temporal profile (logarithmic scale), and (b) spectral profile of the output pulse.}
	\label{fig:figure5}
\end{figure} 
  
\subsection{Discussions}
To understand the pulse dynamics physically, we analyze the results obtained both qualitatively as well as quantitatively. Fibers with a characteristic parameter, soliton number $N$ much higher than 10 are called highly nonlinear fibers. Realization of self-similar PPs over long distances in such highly nonlinear passive optical fibers is restricted by few important factors. $N$ gives a relative measurement of SPM and GVD effect and determines which effect to dominate the pulse evolution. Further, influence of GVD is controlled by $L_D$ which in turn is restricted by $T_0$ and $\beta_2$. Thus, short pulse width and large $\beta_2$ value reduces $L_D$, increasing the dispersion effect. On the other hand, high pulse peak power $P_0$ and large $\gamma$ value increases the nonlinear effect via smaller $L_{NL}$. Moreover, PP formation is accomplished at particular range of chosen $\beta_2$ and $\gamma$ values. Excessive nonlinearity is failed to be balanced by present dispersion which results in separate role of GVD and SPM on pulse evolution leading to OWB. Thus, PPs are formed in such SPM dominated fibers with a much shorter self-similar propagation regime. Besides, dominating GVD again results in pulse distortions, further reshaping etc. leading to absence of any desired self-similair regime. Our proposed scheme enhances the aforesaid self-similar propagation regime by tailoring the longitudinal dispersion profile along with corresponding away from zero rapid nonlinear modulations over long distances. In our designed fiber, F1 is a similar kind of dispersion-decreasing as well as nonlinearity-increasing fiber. Along F1, $L_D$ is varying over a wide range, from a few meters to several kilometers which dictates a gradual decrement of GVD effect, while $L_{NL}$ is in the range of centimeters causing large nonlinearity. $N$ is varying from 7 to 30 for F1. Through F1, PPs have been formed efficiently from high power short input pulse at $\sim$ 1.5 m from the input end of the fiber. In order to stabilize the formed PPs over longer propagation distance, $\beta_2$ is gradually reduced to a value close to zero and allowed to propagate through a rapidly varying longitudinal dispersion profile with nearly-zero average. Simultaneously the nonlinear coefficient has been modulated with a non-zero mean value. For F2-fiber-series, the mean of the rapid $\beta_2$ variation around 0.005 $p$s$^2$/m is maintained over a long propagation distance. Thus, GVD accumulated phase-shifts do not influence the further temporal broadening of the pulse and hence we get a self-similar PP evolution over long ranges. Moreover, the longitudinally modulated nonlinearity of the fiber only broadens the spectrum without destabilizing the temporal profile of the PP. As an evidence of self-similar propagation, the output intensity profile obtained from numerical simulation agrees well with the asymptotic analytical prediction. We have computed the variation of pulse width $\tau$ (normalized) and normalized power $P_n$ along the entire HNBF length and the ratio $\tau$/$P_n$ explicitly dictates a nearly constant variation over length with RSD $\sim$ 2.28$\%$, thus establishing the self-similar scaling. Moreover, at the output end of the designed HNBF, we have achieved a smooth parabolic pulse with a linear chirp across the pulse width. Fig. \ref{fig:figure5}(a) shows the logarithmic plot of normalized power profile. The top-hat nature with steep vertical edges over 2 orders of magnitude is the hall-mark of being the pulse shape essentially parabolic. Further, we have compared the output temporal and spectral profile with deliberate inclusion of loss (Fig. \ref{fig:figure5}(a) and (b) respectively). In this context, we assumed initially only the confinement loss (i.e. 4.516$\times$10$^{-5}$ dB/m) which is almost negligible for our specific fiber design. In order to address the influence of loss due to 6 splices, we have neglected intrinsic losses assuming splicing of two fibers with identical geometries and mode field diameters, and considered only accumulated extrinsic loss of total 0.24 dB (i.e. maximum of 0.04 dB loss for each identical fiber splicing). It is clear from Fig. \ref{fig:figure5} that incorporation of loss only reduces the energy content keeping pulse and spectrum shapes intact over longer propagation length. Moreover, in Fig. \ref{fig:figure4}(b), the chirp evolution is depicted. It is evident that starting from an un-chirped condition, the chirp developed at 2 m of HNBF length from input end has a perfect linear nature without oscillations across the entire pulse width. With further propagation, the linearity of the chirp is constricting to the center portion of the entire pulse duration with a sharp change at the edges. It can be understood as an effect of increasing nonlinear phase-shifts however oscillations have not appeared yet. Moreover, large phase accumulation due to excessive nonlinearity broadens the spectral waveform significantly with large ripples at the top. The proposed rapid dispersion variation along with nonlinearity modulation over length could be otherwise achieved by a single fiber with simultaneous up-and-down tapering. However, fabrication of such complicated structures via state-of-the-art fabrication processes would be more challenging. Hence from the view-point of the feasibility for fabrication, we have designed the segmented fiber consisting of a 1 m long HNBF with zero-crossing longitudinal dispersion profile that can be spliced with another identical fiber in reverse order and the process could be continued over longer distances with a loss compromise.
\section{Conclusion}
In this paper, a novel scheme of rapidly varying longitudinal dispersion profile around a mean value close to zero with simultaneous modulation in nonlinear coefficients, has been proposed in order to achieve self-similar propagation in passive highly nonlinear fiber. To implement the proposed scheme, we have designed a $GeAsSe/AsSe$ glass based highly nonlinear and segmented bandgap fiber. At 2.8 $\mu$m, a self-similar parabolic pulse with 4.12 $p$s FWHM and $\sim$ 38 nm of 3dB spectral broadening over several meters of propagation has been achieved numerically. Moreover, sustained self-similarity of such parabolic pulses with propagation over longer lengths makes them suitable for mid-IR spectroscopy, biological surgeries and imaging. We claim this finding as first ever report of formation and self-similar propagation of such PPs over further distance along a highly nonlinear passive fibers in the mid-IR.
\section*{Acknowledgements}
This work was partially funded by Department of Science and Technology India under INSPIRE Faculty Fellow [IFA-12; PH-23] scheme. BPP acknowledges partial support received from Department of Navy (USA) Grant N62909-10-1-7141 issued by ONRG.

\end{document}